\documentclass[aps,prl,preprint,showpacs]{revtex4}
\usepackage{amssymb}
\usepackage{amsmath}
\usepackage{epsfig}
\usepackage{color}

\newcommand {\cn} {{\rm cn}}

\begin{document}

\title{Centrifugally driven relativistic particles: general treatment 
and new solutions}
\author {Giorgi Khomeriki}
\affiliation {Centre for Theoretical Astrophysics, ITP, Ilia State University, 0162-Tbilisi, Georgia}
\affiliation {Physics Department, Tbilisi State University, 3 Chavchavadze, 0128 Tbilisi, Georgia}
\affiliation {Vekua school of Physics \& Mathematics, 9 Tchaikovsky, 0105 Tbilisi, Georgia}

\author {Andria Rogava}
\affiliation {Centre for Theoretical Astrophysics, ITP, Ilia State University, 0162-Tbilisi, Georgia.}

\begin{abstract}
Dynamics of a centrifugally driven particle moving along a magnetically prescribed trajectory (e.g. pulsar magnetic field line) in a spacetime of a rotating massive body is considered. We do assume that the particle is on a zero Landau level and is not losing energy via synchrotron or any other kind of radiation. Based on this model we further study in detail the zero-gravity limit, reconsider the "straight field-line" {\it gedanken experiment} case \cite{mac94} in the "bead-on-the-wire" approximation and find previously overlooked exact analytic solutions for the most general set of initial values. We analyze different regimes of a particle  dynamics, classifying them in their relation with specific initial conditions. We conclude by indicating further directions of the study and discussing possible areas of astrophysical applications. 
\end{abstract}

\maketitle

\section{Introduction}

In 1994 Machabeli \& Rogava described  \cite{mac94} an elemental {\it gedanken experiment}: centrifugally driven motion of a bead inside a long, straight pipe rotating with a constant angular velocity $\omega = const$ around an axis normal to its symmetry axis. A particular initial condition has been selected: at $t=0$ the bead was supposed to be at $r_0 =0$, right above the pivot, possessing an initial velocity $v_0$. An analytic solution was found,  exhibiting exotic regimes of the particle dynamics: namely, it was revealed that centrifugally driven particles could both accelerate and decelerate, their acceleration could change its sign; actual modes of motion strongly depended on the value of $v_0$. The importance of this result in the context of centrifugally driven dynamics of particles in rapidly rotating pulsar magnetospheres and astrophysical jets was later repeatedly pointed out \cite{rog03} and extensively studied in a number of subsequent publications \cite{gan96, osm07, gud15, mac16, osm17}. In particular, particle acceleration by rotating magnetospheres in active galactic nuclei was investigated \cite{rig00}; the role of radiation reaction forces in the dynamics of centrifugally accelerated particles was explored \cite{dal07}; centrifugal acceleration was studied in isotropic photon fields \cite{bak17, bak18} and wormhole metrics \cite{ars17};

In ref  \cite{mac94} a bead is considered to be inside very long pipe, which can rotate around its perpendicular axis (for simplicity we also consider rotation and magnetic field axis to be perpendicular). 
The bead is given initial speed very near to rotational axis and it turns out that when relativistic effect is considered, the bead performs oscillatory motion. Upon reaching the light cylinder,
the radial velocity of the bead becomes zero and it starts to move towards rotational axis. The main peculiarity of the motion is that centrifugal force is directed towards the rotational axis.
 Of course, this model is partially unrealistic because of the fact that bead will never be able to oscillate, as no pipe is strong enough to hold a particle with almost infinite energy, 
but the approach and results still are spectacular as the origin of negative centrifugal force is manifested in simple relativistic system. 

We associate this simple mechanical-relativistic problem with Pulsars. Pulsar's magnetic field is very strong and the effect of "freezing" particle to magnetic field lines is possible. Every particle that enters Pulsar's magnetic field, either from surface or from accretion disk, starts cyclonic motion around field line and immediately (after some pico seconds) goes to quantum (Landau) levels. This effect is called "freezing" to field lines.
Because of this freezing, many scientists consider bead-rod problems. However, in the very first paper \cite{mac94} that used this approach, authors only consider a particle to be at rotation axis and have some initial speed (which they vary and observe the results). In real astrophysical situations, usually a particle is influenced by Pulsar magnetic field when it leaves a thick layer of plasma on the surface. So, in real astrophysical scenarios the conditions of the relativistic bead-rod problem are as follows: initial distance from the rotational axis is not zero, but approximately the radius of Pulsar. This might be negligible for Normal Pulsars as their radius is several kilometers and the radius of light cylinder is $3\cdot10^5$ km (if we assume that angular frequency is $\omega=1$). But for example as for millisecond Pulsars ($\omega$ can be up to 3000) the light cylinder is much closer and this distance becomes comparable to Pulsar's radius.  
Another possibility is that if we have a binary system and accretion disk is formed around Pulsar, particles can start "falling" on Pulsar surface. In this case one might get a scenario when a particle will not be able to reach the surface because it will get decelerated by centrifugal force, etc.

\section{Main Consideration}

The purpose of this study is to consider motion of centrifugally driven relativistic particles along prescribed trajectories  embedded in a general-relativistic space-time of a rotating central object. In particular, the particles are constrained to move along filed lines by the "frozen-in" condition. In the next subsection we present the generalization of the formalism which was partly developed in \cite{rog03} for flat prescribed curved trajectories and in \cite{gud15} for prescribed trajectories within gravitational fields generated by rotating bodies.  

\subsection{General formalism}

A rotating massive object, for instance, a neutron star or a black hole, is described by a non-diagonal spacetime metric, with its elements independent of $t$ and $\phi$ and having the following form \cite{str84}:
\begin{equation}
\label{metr} ds^2 = - d \tau^2 =
g_{tt}dt^2+2g_{t\phi}dtd\phi+ g_{\phi\phi}d\phi^2 + g_{r r}dr^2 + g_{\theta \theta}d\theta^2,
\end{equation}

In the nonrelativistic limit this metric reduces to the
Minkowski metric, written in spherical coordinates. In the presence of gravity but the absence/presence of the central body rotation it
reduces to non-rotating (Schwarzschild or Reissner-Nordstrom) and rotating (Kerr or Kerr-Newman) black hole metrics, respectively. In a geometrized unit system, where $G =  c = 1$, the Kerr metric, for example, is written as \cite{sha83}:
\begin{equation}
g_{tt} = - \left[ 1 - {2Mr}/{\Sigma} \right]
\end{equation}
\begin{equation}
g_{t \phi} = - {2aMr sin^2 \theta}/{\Sigma}
\end{equation}
\begin{equation}
g_{\phi \phi} = \left[ r^2 + a^2 + 
{2M a^2 r sin^2 \theta}/{\Sigma}  \right]
 sin^2 \theta
\end{equation}
\begin{equation}
g_{rr} = {\Sigma}/{\Delta}
\end{equation}
\begin{equation}
g_{\theta \theta} = \Sigma
\end{equation}
where $\Sigma \equiv r^2 + a^2 cos^2 \theta$, $\Delta \equiv r^2 - 2Mr + a^2$, while $a \equiv J/M$  and $J$ and $M$ are the angular momentum and the mass of the central object, respectively. One can easily check that When $a=0$ it obviously reduces to the case of nonrotating black hole (Schwarzschild metric) and if further $M/r <<1$ it reduces to Minkowksi metric written in (asymptotically) spherical $[r, \theta, \phi]$ coordinates.

In different astrophysical situations plasma particles may have various kinematically complex motions involving ejection and/or rotational motion within the considered spacetime. But as we have already noted strong magnetic fields constrain plasma particles to move along the field lines. Then the notion of ``prescribed trajectories'' \cite{rog03} turns out to be quite useful. It naturally leads to the idea to "embed" the trajectories, predefined by the geometry of magnetic field, within the texture of the metric itself. It follows that the study of the dynamics of particles, moving along these prescribed ``rails'' (field-lines) is simplified by the fact that the shape of particle trajectories coincides with that of field lines and is known in advance. This important circumstance comprises the core of the  "bead-on-the-wire" approximation \cite{mac94, gan96, rog03}. Physically this model presupposes that the magnetic field energy density exceeds that of the plasma kinetic energy by many orders of magnitude.

Let us consider differentially rotating set of trajectories, specified by the angular velocity $\Omega(r)$ and make the following transformation of variables in the (1) metric:
\begin{equation}
\theta = \theta(r), ~~~  \varphi = \varphi (r) + \Omega(r)t.
\end{equation}

It will lead us to the metric of the rotating frame of reference, for the prescribed trajectories, which can readily be written in the following way:
\begin{equation}
\label{met} ds^2 = {\cal G}_{tt}dt^2 + 2{\cal G}_{tr}dtdr + {\cal G}_{rr}dr^2,
\end{equation}
where
\begin{equation}
{\cal G}_{tt} \equiv g_{tt} + 2 \Omega g_{t \varphi} + \Omega^2 g_{\varphi \varphi}
\end{equation}
\begin{equation}
{\cal G}_{t \varphi} \equiv (\varphi' + \Omega't)(g_{t \varphi} + \Omega g_{\varphi \varphi})
\end{equation}
\begin{equation}
{\cal G}_{rr} \equiv g_{rr} + \theta'^2 g_{\theta \theta} + (\varphi' + \Omega't)^2 g_{\varphi \varphi}
\end{equation}
Note that in these equations primes denote radial derivatives, i.e., $\varphi' \equiv d \varphi /dr$, $\Omega' \equiv d \Omega /dr$ and $\theta' \equiv d \theta /dr$. Note also that the ${\cal G}_{ik}$ metric is not stationary anymore but has explicit time-dependence arising due to the presence of differential rotation. It is evident that when this is the case proper energy of a particle moving on such a prescribed trajectory wouldn't be a conserved quantity!

It is easy to see that the dynamics of a particle moving along the prescribed trajectory can be defined in terms of the Lagrangian $2 {\cal L} \equiv {\cal G}_{\alpha \beta} \dot{x}^{\alpha} \dot{x}^{\beta}$ and the subsequent equations of motion:
\begin{equation}
\label{equat} \frac{\partial {\cal L}}{\partial x^{\alpha}} =
\frac{d}{d\tau}\left(\frac{\partial {\cal L}}{\partial
\dot{x}^{\alpha}}\right),
\end{equation}
where $u^\alpha \equiv \dot{x}^{\alpha} \equiv dx^{\alpha}/{d\tau}$, and $x^{0}\equiv t,\;\;\; x^{1}\equiv r$.

For the t-component of this equation, for instance, we will have:
\begin{equation}
\frac{d}{d \tau}{\left[{\cal G}_{tr} u^r + {\cal G}_{tt} u^t  \right]} = \Omega'
{\left[(g_{t \varphi} + \Omega g_{\varphi \varphi}) u^t u^r + (\varphi' + \Omega't)
g_{\varphi \varphi} u^r u^r \right]}
\end{equation}

This equation shows what we have already envisaged: the presence of the differential rotation makes a particle's proper energy non-conserved quantity and significantly complicates the mathematical description of the particle motion. In this paper we will limit our study with the consideration of rigidly rotating trajectories. Hence, hereafter we will assume that $\Omega' =0$. In this case we can define conserved proper energy as:
\begin{equation}
\label{energy} E = - u_t = -u^t \left({\cal G}_{tr} v + {\cal G}_{tt}\right) = const,
\end{equation}
while the four-velocity normalization condition ${\cal G}_{\alpha \beta} u^\alpha u^\beta = -1$ gives:
\begin{equation}
u^t u^t ({\cal G}_{tt} + 2v {\cal G}_{tr} + v^2 {\cal G}_{rr}) = -1,
\end{equation}
where the radial velocity $v \equiv dr/dt$. From these two equations we can derive an explicit quadratic equation for the radial velocity:
\begin{equation}
a v^2 + b v + c = 0,
\end{equation}
where
\begin{equation}
a \equiv {\cal G}_{rr} E^2 + {\cal G}_{tr}^2,
\end{equation}
\begin{equation}
b \equiv 2 {\cal G}_{tr} (E^2 + {\cal G}_{tt}),
\end{equation}
\begin{equation}
c \equiv {\cal G}_{tt} (E^2 + {\cal G}_{tt}).
\end{equation}

Formal solution of this equation is:
\begin{equation}
v = \frac{1}{2a}(-b \pm \sqrt{D}),
\end{equation}
where
\begin{equation}
D \equiv 4 E^2 ({\cal G}_{tr}^2 - {\cal G}_{tt} {\cal G}_{rr})
(E^2 + {\cal G}_{tt}).
\end{equation}

With the general treatment highlighted in this section, one can solve many interesting problems of a centrifugally driven relativistic particles on various prescribed trajectories in different metrics (and thus, near different astrophysical objects).  

\subsection{Special-relativistic case: exact solution}

In this section, we will get new solution for the simplest case of this kind of problem. 
Let us consider the motion in normal Minkowskian metric (in units of $G=c=1$):
\begin{eqnarray}
{ds^2}=-{d\tau^2}=-{dT^2}+{dX^2}+{dY^2}
\label{intervali}
\end{eqnarray}
Making the following transformation of variables (for simplicity, also considering that rotational axis is perpendicular to XY plane):
\begin{eqnarray}
T=t \quad X=rcos\left(\omega t\right) \quad Y=rsin\left(\omega t\right)
\end{eqnarray}
The two-dimensional metric in the rotating frame will be the following:
\begin{eqnarray}
-{d\tau^2}=-\left(1-{\omega^2}{r^2}\right){dt^2}+{dr^2}
\label{metrica}
\end{eqnarray}
where $\omega$ is rotational angular frequency and r is distance from the rotational axis. Thus the Lagrangian reads as:
 \begin{eqnarray}
L=\frac{1}{2}\left[-\left(1-{\omega^2}{r^2}\right)\left(\frac{dt}{d\tau}\right)^2+\left({\frac{dr}{d\tau}}\right)^2 \right]
\label{lagran}
\end{eqnarray}
From \eqref{lagran},  we can write Lagrange equation:
 \begin{eqnarray}
\frac{d}{d\tau}\left(\frac{\partial L}{\partial \dot{x}^\alpha}\right)=\frac{\partial L}{\partial x^\alpha}
\end{eqnarray}
And given Lagrange equation can be written:
 \begin{eqnarray}
-\left(1-{\omega^2}{r^2}\right)\left(\frac{dt}{d\tau}\right)=const \equiv -E
\label{invariant}
\end{eqnarray}
 \begin{eqnarray}
\frac{d^2 r}{d\tau^2}={\omega^2}r\left(\frac{dt}{d\tau}\right)^2
\label{motion1}
\end{eqnarray}
From \eqref{invariant},  \eqref{motion1} and also by using the metric \eqref{metrica} we can get governing equation of motion for the bead.

\begin{eqnarray}
\frac{dr}{dt}=\sqrt{\left(1-{\omega^2}r^2\right)\left(1-\frac{\left(1-{\omega^2}r^2\right)}{E}\right)}
\label{motion2}
\end{eqnarray}
where E is invariant and only depends on initial conditions as follows: according to \eqref{motion2}, at $t=0$, let us denote that $\frac{dr}{dt}=v_0$ and $r=r_0$, thus E will be:
\begin{eqnarray}
E=\frac{1-{\omega^2}r_0^2}{\sqrt{1-{\omega^2}r_0^2-{v_0}^2}}
\label{invarianti}
\end{eqnarray}

The character of the solution of this equation depends on the value of E. If $E>1$ the solution will be elliptic cosine that was derived originally in Ref. \cite{mac94}. If $E<1$ solution will be different (but see below).
In order to find the solutions of equation \eqref{motion2}, we should define the variables $\theta$, $\lambda$ and $m$ in the following way:
\begin{eqnarray}
\theta=acos\left(\omega r\right)   \quad T=\omega t \quad m=\frac{1}{E^2}
\end{eqnarray}
And get a simplified equation:
\begin{eqnarray}
\frac{d\theta}{dT}=-\sqrt{1-m{{\sin^2{\theta}}}}
\end{eqnarray}
Which is the definition of elliptic cosine (cn function), if $m<1$ and thus:
\begin{eqnarray}
r(t)=\frac{1}{\omega}\cn\left(\lambda-{\omega}t,m\right)
\label{cn}
\end{eqnarray}
where  $\lambda$ is incomplete elliptic integral of the first kind:.
\begin{eqnarray}
{\lambda}=\int_{0}^{\phi}{\frac{d\theta}{\sqrt{1-m\sin^2{\theta}}}}
\label{integral}
\end{eqnarray}
where $\phi=acos(\omega r)$, $m=1/E^2$ note that m should be less than 1, otherwise given integral will be undefined. 

from equation \eqref{cn}, it is clear that the motion will be oscillatory, the particle will have zero radial speed when it reaches the light cylinder and it will start moving towards rotation axis after that point.
In real systems, this oscillations will not happen, because the magnetic field lines will not be able to hold 
a particle with infinite energy and thus its tangential velocity $\omega$$r$ will not reach the speed of light. Before reaching light cylinder,
 particles will start bending magnetic field lines and at some point they will not be influenced by them at all.

As we mentioned earlier, \eqref{cn} does not describe the motion 
if initial energy $E<1$ (m is proportional to $\frac{1}{E^2}$ , so when $E<1$, elliptical integral according to eq. \eqref{integral} is undefined).  

In order to get analytical solutions for Eq. \eqref{motion2}, in case when $E>1$, we  denote $m=E^2$ and the solution for Eq. \eqref{motion2} reads as:
\begin{eqnarray}
r(t)=\frac{1}{\omega}dn\left(\frac{\omega}{E}t-\lambda,m\right)
\label{dn}
\end{eqnarray}
In this case, $\lambda$ is the same as in Eq. \eqref{integral}, but with different $\phi=asin\sqrt{\frac{1-{\omega^2}r_0^2}{m}}$ and  $m=E^2$ .

If we look at invariant Energy which only depends on initial conditions in Eq. \eqref{invarianti}, we can physically determine which analytical solution (Eq. \eqref{cn} or Eq.\eqref{dn}) should be used. If initial radial velocity is zero, energy (E) will always be more than one (unless $r_0=0$, which will mean that particle is on the rotational axis and does not have any velocity. This is unstable trivial  solution), so the solution for this initial conditions will be Eq. \eqref{dn}. On the other hand, if particle is initially located on the rotational axis ($r_0=0$), the solution will be Eq. \eqref{cn}.

And in real physical case, when particle starts its cetrifugal motion with some given velocity and also the distance to rotational axis is not negligible, only one from  Eqs. \eqref{cn} and \eqref{dn} will be the solution to this original problem.

\begin{figure}[!h]
		\includegraphics[scale=0.6]{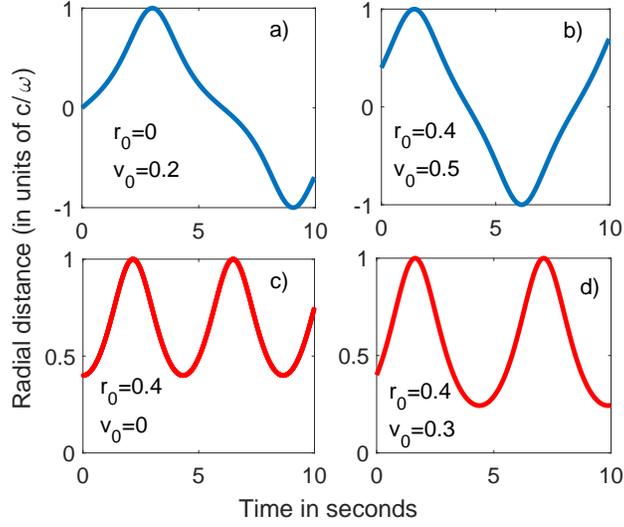}
		\caption{These figures represent dependance of particle's distance from rotational axis on time. a) and b) are graphs of cn function. c). and d) are for dn function. Initial conditions is given on each graph, and for all cases Rotator's (Pulsar's) rotational angular frequency is $\omega=1$}
\end{figure}

\begin{figure}[!h]
		\includegraphics[scale=0.6]{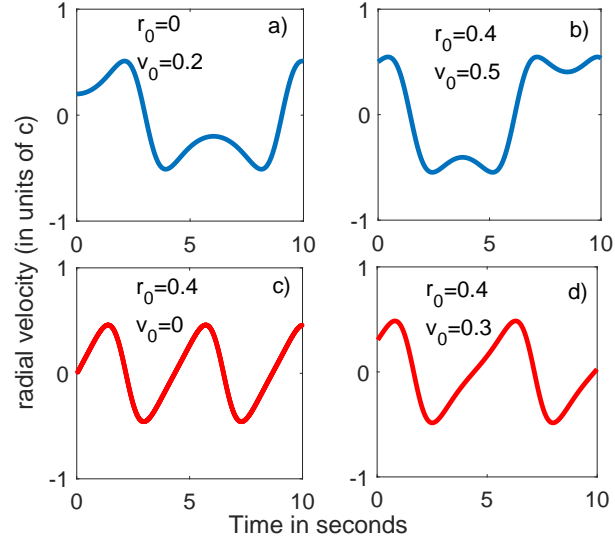}
		\caption{These figures represent dependence of particle's absolute velocity on time (we have plotted absolute velocity because the orientation of particle's motion is easily determinable from both these and previous graphs) . a) and b) are graphs of cn function. c). and d) are for dn function. Initial conditions is given on each graph (and are the same as in the previous figure), and for all cases Rotator's (Pulsar's) rotational angular frequency is $\omega=1$}
\end{figure}

\begin{figure}[!h]
		\includegraphics[scale=0.6]{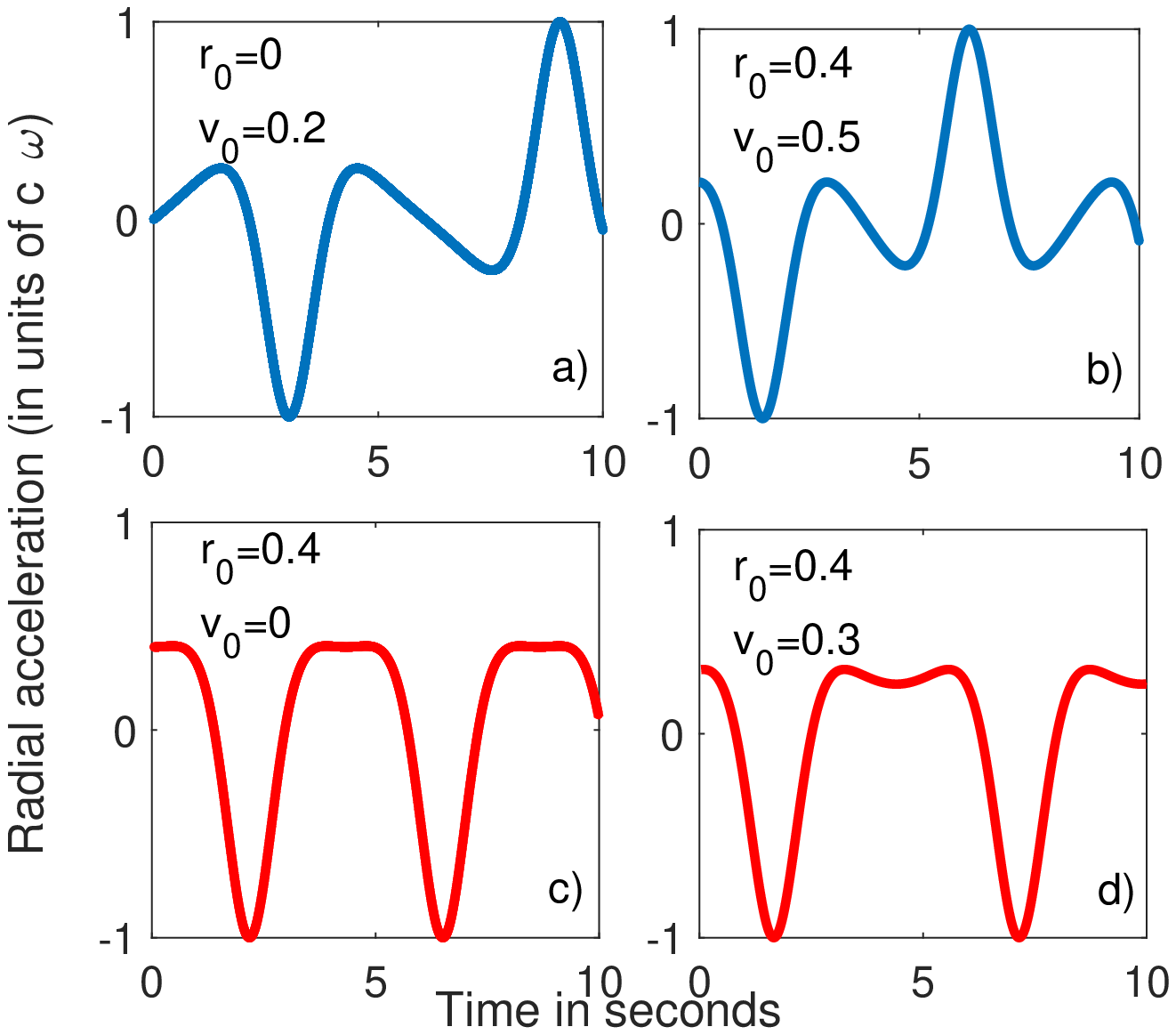}
		\caption{These figures represent dependence of acceleration on time. a) and b) are graphs of cn function. c). and d) are for dn function. Initial conditions is given on each graph (and are the same as in previous figures), and for all cases Rotator's (Pulsar's) rotational angular frequency is $\omega=1$}
\end{figure}

\noindent{\it Results:}
After deriving generalized equations of motion for any initial parameters (elliptic functions cn and dn), now we will study particle's behavior for different initial conditions.

Figures 1. 2. and 3.  show us respectively the graphs (for $\omega=1$) of radial distance dependence on time, radial velocity dependence on time and radial acceleration dependence on time. In every Figure a) and b) subfigures represent solutions for elliptic cosine. c) and d) subfigures represent solutions for elliptic function dn.

In FIG. 1 one can see particle's radial distance (from rotational axis) dependence on time.  subfigure a) shows that the solution will be elliptic cosine if in the beginning particle is on the rotational axis and has some initial velocity, and reversely, if initially particle is far from rotational axis and has velocity equal to zero (subfigure c ), its motion will be elliptic function dn. However, if initially  distance from rotational axis and velocity both are nonzero, the solution will be either cn or dn. if velocity is large enough for the particle to reach rotational axis with nonzero speed, then the solution will be cn (subfigure b), and otherwise (if the particle will not be able to reach rotational axis and will stop somewhere) the solution will be dn. Also, we can distinguish these two motions (cn and dn) easily: it is cn, if particle oscillates in the range between -1 and 1. it is dn, if particle oscillates in the range between 0 and 1 (the upper boundary should be 1. However, lower limit is not fixed and can be anything from 0 to 1, it depends on initial conditions).

Figure 2. shows particle's absolute radial velocity dependence on time for the same initial conditions as in FIG. 1. 

One of the main peculiarities of this problem is that acceleration is not always positive and particle can be attracted towards rotational axis (negative value of acceleration is only at relativistic velocities). In Fig. 3 one can observe  particle's acceleration dependence on time, for the same parameters as in Fig. 1. We should also note that acceleration can even be negative at the very beginning of motion, if the initial kinetic energy is too large. For example, if $r_0=0$ and $v_0>\frac{sqrt{2}}{2}$, acceleration will be negative at $t=0$. However, if $v_0=0$ acceleration can not be negative for any initial distance $r_0$.

All the figures have been plotted for $\omega=1$ (rotational angular frequency). We are suggesting that in many real physical systems, such as millisecond Pulsars, one would need to use the solution of elliptic function dn, because millisecond Pulsars have very small Period (thus very large angular frequency, up to 1000-3000 Hz), and even Pulsar's radius is comparable to the length of its light cylinder.  Thus, questions may arise why we only plotted graphs for $\omega=1$: That is because, changing $\omega$ does not change the characteristics of particle's motion. The shape of graphs remain the same, only change is in time needed to reach the light cylinder, the absolute values of velocities and accelerations, but the shape of the Plot is the same and there is nothing interesting except numbers.

Also, we did not Plot the dependence of Lorentz factor on time or any other parameter. That is because of the fact that during the motion, Lorentz factor increases and tends to infinity near light cylinder for any initial parameter.  That is why in real physical systems no magnetic field will be able to hold the particle on Landau levels near light cylinder. As particle's mass grows, it will be less and less affected by magnetic field and after some point the particle will be moving linearly. 

\section{Discussion and Conclusion}

We have mentioned many times that this is the idealized problem, not because of the fact that we consider particle "frozen" to the magnetic field lines (this is very realistic for very strong magnetic fields, such as near Pulsars, particles are on Landau levels and are not able to emit any synchrotron radiation), but because at light cylinder particle's total velocity becomes equal to speed of light (not for finite time) and thus the mass increases rapidly as the particle gets closer to the light cylinder. So, magnetic field itself has to be infinitely large to hold a particle on Landau levels and accelerate it centrifugally. 
So, we only say that for realistic systems our equations are only valid before the particle gets very close to light cylinder. 

In fact, to consider a problem of particle's behavior near light cylinder would be very interesting (and we plan to do it in our future works). Many behaviors may occur, for example particle (which will be very massive) can "escape" from Landau levels and emit synchrotron radiation, or we might find a point from where on magnetic field's influence might become negligible and the motion could become linear. 

One can also consider radiation (because of accelerated motion) from particle's centrifugal motion described in this work and compare its spectra to real physical objects.

In conclusion, we would like to say that we have found another very important solution for this original problem and alongside with highlighting the results we discussed in which physical scenarios one can use the solution that is represented by elliptical function dn. In particular, it can be used to describe particle's motion alongside millisecond Pulsar's straight magnetic field lines, because Pulsar radius is comparable to light cylinder diameter and in many cases (it depends on initial velocity) second solution (elliptical function dn) would be needed to describe the motion.

\section{Acknowledgements}
Both authors are grateful to the Centre for Mathematical Plasma-Astrophysics, KU Leuven, where the final part of this study was performed. Our research was supported by Shota Rustaveli National Science Foundation of Georgia (SRNSFG), grants FR-18-564 and FR-18-14747.

\end{document}